\documentclass{article}

\usepackage[preprint]{neurips_2023}

\usepackage[utf8]{inputenc} 
\usepackage[T1]{fontenc}    
\usepackage{hyperref}       
\usepackage{url}            
\usepackage{booktabs}       
\usepackage{amsfonts}       
\usepackage{nicefrac}       
\usepackage{microtype}      
\usepackage{xcolor}         

\usepackage{multirow}
\usepackage{color}
\usepackage{colortbl}
\usepackage{stfloats}
\usepackage{comment}
\usepackage{proof}
\usepackage{fancybox}
\usepackage[flushleft]{threeparttable}
\usepackage{wrapfig}
\usepackage{pgfplotstable} 
\pgfplotsset{compat=1.12}
\usepackage{tabularx}
\usepackage{arydshln}
\usepackage{hyperref}
\usepackage{tcolorbox}
\usepackage{subfig}
\usepackage[noend]{algpseudocode} 

\usepackage{bm}
\usepackage{mathtools}
\usepackage{stmaryrd}
\usepackage{graphicx}
\usepackage{color}
\usepackage{colortbl}
\usepackage{stfloats}
\usepackage{comment}
\usepackage{proof}
\usepackage{fancybox}
\usepackage[flushleft]{threeparttable}

\usepackage{paralist, tabularx}
\usepackage{wrapfig}

\usepackage{pgfplotstable} 
\pgfplotsset{compat=1.12}

\usepackage{tabularx}

\usepackage{arydshln}
\usepackage{hyperref}

\definecolor{lightgray}{HTML}{eeeeee}
\definecolor{mildgray}{HTML}{D6D6D6}
\definecolor{darkgray}{HTML}{B5B4B5}

\newcommand{\tbmgray}{\cellcolor{mildgray}}

\newcommand{\website}{\url{https://sites.google.com/view/ai-cps-robotics-manipulation/home}}

\title{Towards Building AI-CPS with NVIDIA Isaac Sim:
An Industrial Benchmark and Case Study for Robotics Manipulation}

\author{%
  Zhehua~Zhou\thanks{Contributed equally to this research.} \\\
  University of Alberta\\
  \texttt{zhehua1@ualberta.ca} \\
  \And
  Jiayang~Song$^*$ \\\
  University of Alberta\\
  \texttt{jiayan13@ualberta.ca} \\
  \And
  Xuan~Xie$^*$ \\\
  University of Alberta\\
  \texttt{xxie9@ualberta.ca} \\
  \And
  Zhan~Shu \\\
  University of Alberta\\
  \texttt{zshu1@ualberta.ca} \\
  \And
  Lei~Ma \\\
  The University of Tokyo \\
  University of Alberta\\
  \texttt{ma.lei@acm.org} \\
  \And
  Dikai~Liu \\\
  NVIDIA AI Tech Centre\\
  \texttt{dikail@nvidia.com} \\
  \And
  Jianxiong~Yin \\\
  NVIDIA AI Tech Centre\\
  \texttt{jianxiongy@nvidia.com} \\
  \And
  Simon~See \\\
  NVIDIA AI Tech Centre\\
  \texttt{ssee@nvidia.com} \\
}

\begin{document}

\maketitle

\begin{abstract}
As a representative cyber-physical system (CPS), robotic manipulator has been widely adopted in various academic research and industrial processes, indicating its potential to act as a universal interface between the cyber and the physical worlds.
Recent studies in robotics manipulation have started employing artificial intelligence (AI) approaches as controllers to achieve better adaptability and performance.
However, the inherent challenge of explaining AI components introduces uncertainty and unreliability to these AI-enabled robotics systems, necessitating a reliable development platform for system design and performance assessment.
As a foundational step towards building reliable AI-enabled robotics systems, we propose a public industrial benchmark for robotics manipulation in this paper.
It leverages NVIDIA Omniverse Isaac Sim as the simulation platform, encompassing eight representative manipulation tasks and multiple AI software controllers.
An extensive evaluation is conducted to analyze the performance of AI controllers in solving robotics manipulation tasks, enabling a thorough understanding of their effectiveness.
To further demonstrate the applicability of our benchmark, we develop a falsification framework that is compatible with physical simulators and OpenAI Gym environments. 
This framework bridges the gap between traditional testing methods and modern physics engine-based simulations.
The effectiveness of different optimization methods in falsifying AI-enabled robotics manipulation with physical simulators is examined via a falsification test. 
Our work not only establishes a foundation for the design and development of AI-enabled robotics systems but also provides practical experience and guidance to practitioners in this field, promoting further research in this critical academic and industrial domain.
The benchmarks, source code, and detailed evaluation results are available on \website.
\end{abstract}

\section{Introduction}
\label{sec.intro}
Over the past decades, cyber-physical systems (CPSs) have emerged as an important area of research and development across various industrial sectors \cite{jazdi2014cyber,wolf2009cyber,chen2017applications,Song2021WhenCS}.
As a complex system that integrates physical processes with software elements, modern CPS often comprises a diverse range of applications, e.g., autonomous vehicles~\cite{schwarting2018planning,kato2015open}, smart grids~\cite{tuballa2016review,ma2013smart}, and medical devices~\cite{dey2018medical}. 
Among these applications, robotic manipulator holds great importance due to its wide-ranging industrial applications, spanning domains such as manufacturing~\cite{billard2019trends}, logistics~\cite{mason2018toward}, healthcare~\cite{kolpashchikov2022robotics}, and agriculture~\cite{zhang2020state}.
Serving as a potential universal interface bridging the software and the physical world, robotic manipulator enables in-depth cyber-physical interactions. 
Recent studies in robotics manipulation have extensively embraced artificial intelligence (AI) techniques, particularly deep reinforcement learning (DRL) methods, as controllers to overcome the challenges and limitations associated with traditional control software~\cite{karoly2020deep}.
As a notable example of AI-enabled CPS (AI-CPS)~\cite{radanliev2021artificial}, these AI-enabled robotics manipulation systems exhibit the potential to address a broad spectrum of complex tasks and effectively handle changing environments, thereby opening up new avenues for achieving higher levels of autonomy and universality.


However, the integration of AI techniques into robotics systems also introduces uncertainty and unreliability due to the challenges in interpreting the behavior of AI software controllers~\cite{luo2020balance}, necessitating a powerful development platform for system design and analysis.
Considering factors such as safety, cost, and time efficiency, a reliable and accurate simulator is often the central component of such a development platform~\cite{hofer2021sim2real,hofer2020perspectives}.
In recent robotics research, software simulators based on physics engines have gained preference for simulating complex industrial systems due to their enhanced accuracy, scalability, and flexibility~\cite{werling2021fast, collins2021review}.
To address this need, our industrial partner, NVIDIA, has recently introduced NVIDIA Omniverse Isaac Sim~\cite{isaacsim, makoviychuk2021isaac} (referred to as Isaac Sim in this paper), an advanced physical simulator that offers seamless integration with NVIDIA hardware and software, providing robust support for high-speed, GPU-accelerated simulation and AI training.
Isaac Sim is capable of providing highly realistic simulations that accurately model the system behaviors in the real world, making it an ideal simulator for the development of AI-enabled robotics applications.

\begin{figure*}
\centering
\includegraphics[width=\linewidth]{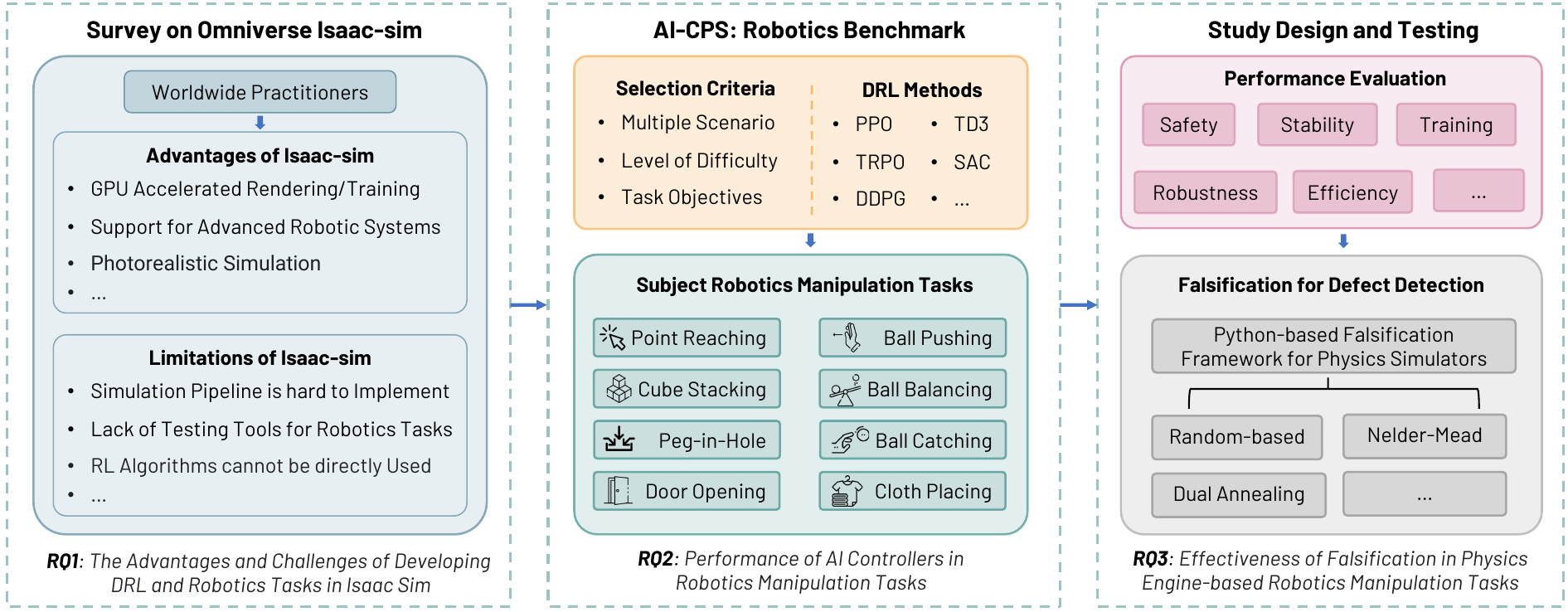}
\caption{Workflow summary of this paper.}
\vspace{-10pt}
\label{fig_workflow}
\end{figure*}

Despite the promising capabilities of Isaac Sim, there remains a need for a deep understanding of its performance in various robotics tasks involving AI components. 
Furthermore, to the best of our knowledge, there is a lack of research on how to effectively and systematically utilize Isaac Sim as a development platform for AI-enabled robotic manipulators.
Therefore, in this paper, we take the first step in investigating the perspectives of practitioners and establishing the foundations to support research and development in AI-enabled robotics manipulation with Isaac Sim.
Fig.~\ref{fig_workflow} shows a high-level overview of our workflow. 
We arrange our investigation and study design with the following steps.

To identify the most critical industrial demands, we first conducted an survey involving academic and industrial practitioners from the global robotics and AI communities. 
Through a series of questions, we gathered valuable insights into the advantages and drawbacks of Isaac Sim compared to other physical simulators. 
In addition, we sought participants' input on their requirements for designing and developing AI-enabled robotics applications.
This survey not only reveals under which application scenarios Isaac Sim may outperform other simulators but also uncovers the community demands and future directions for developing a high-performing physical simulator.
Based on the survey results, we recognize the urgent need for a benchmark to unveil the characteristics of Isaac Sim in the context of developing AI-enabled robotics manipulation.


To bridge this gap, we initiate an early exploratory study that establishes a public industrial benchmark comprising eight representative robotics manipulation tasks, along with multiple software controllers trained with DRL algorithms.
This benchmark behaves as a cornerstone for supporting the design and development of AI-enabled robotics manipulation, promoting further research in this critical academic and industrial domain.
Furthermore, together with our industrial partners, we design the benchmark with great emphasis on extensibility and applicability, allowing for an end-to-end software development lifecycle towards trustworthy AI-enabled systems.
To further evaluate the performance of AI controllers in manipulation tasks simulated with Isaac Sim, we conduct an extensive and in-depth evaluation to assess their capabilities across different tasks.
Our experimental results demonstrate that these controllers deliver satisfactory performance in a wide range of tasks and exhibit a commendable level of robustness against action noise.


In addition, we notice a lack of dedicated testing tools specifically designed for physical simulators like Isaac Sim.
Testing plays a crucial role in the software development lifecycle as it helps identify scenarios or conditions in which the system fails or exhibits unsafe behaviors.
Nevertheless, it remains uncertain whether traditional testing methods are still effective when applied to the scope of physical simulators.
To address this gap and demonstrate the practical applicability of our benchmark, we develop the first Python-based falsification framework that is compatible with physical simulators and OpenAI Gym environments.
Our findings indicate that the effectiveness of traditional falsification methods varies across different tasks. 
Therefore, new testing methods that can combine the characteristics of the task, as well as the information from AI components, are anticipated.

The contributions of this work are summarized as follows.

\begin{compactitem}[$\bullet$]

    \item We conduct a survey among worldwide industrial and academic practitioners to compare Isaac Sim with other state-of-the-art physical simulators. 
    The results offer valuable insights into the strengths and limitations of Isaac Sim and reveal the community demands and potential future developments.
    
    \item 
    We establish a public industrial benchmark of robotics manipulation with AI software controllers.
    Leveraging the capabilities of Isaac Sim, we integrate diverse manipulation tasks with unique challenges into a unified benchmark. 
    It lays the foundation for the development of reliable and adaptable AI-enabled robotics manipulation systems.
    
    \item We perform a large-scale analysis and industrial study to evaluate and explore the performance and characteristics of AI software controllers in robotics manipulation. 
    Such an analysis discloses the benefits and challenges of AI controllers and provides insights into developing robust and competent CPSs.

    \item We develop the first Python-based falsification framework that is compatible with physical simulators and OpenAI Gym environments.
    It bridges the gap between traditional falsification methods and modern simulation platforms.
    We additionally investigate the performance and effectiveness of existing optimization methods in falsifying AI-enabled robotics manipulation tasks simulated using physics engines. 
\end{compactitem}



As an early step towards developing resilient, secure, and fault-tolerant AI-enabled industrial-level robotics systems, our benchmark could potentially serve as a valuable resource for efficiently utilizing Isaac Sim as the development platform for AI software controllers in robotics manipulations, offering practical experience and guidance for researchers and practitioners in this field.
Furthermore, the proposed benchmark is highly extensible and has the potential to support the entire software development lifecycle, encompassing planning, design, testing, and deployment stages. 
The early exploration, benchmark, and study in this emerging direction could potentially contribute to advancements in AI-enabled robotics and CPSs and paves the way for innovative software solutions in various real-world scenarios.

\section{Background}
In this section, we first provide a succinct overview of AI-CPS and AI-enabled robotics manipulation. 
Then, a short introduction to signal temporal logic (STL) and falsification technique, which is an important testing method for CPSs, is presented. 


\begin{figure}[!tb]
    \centering
    \includegraphics[width=0.7\columnwidth]{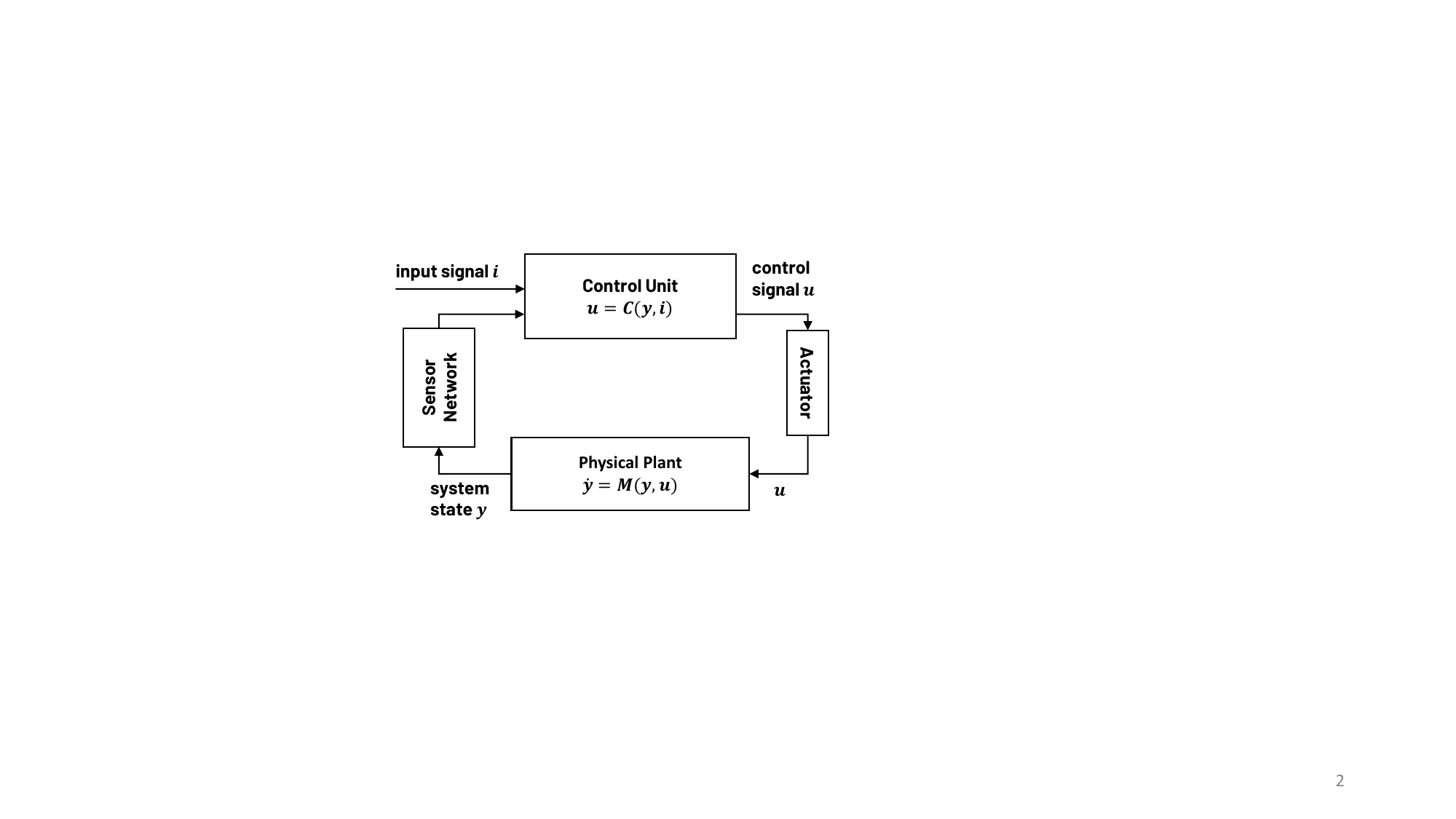}
    \caption{Abstract workflow of CPS}
    \label{fig:CPS_model}
    \vspace{-10pt}
\end{figure}

\subsection{Robotics Manipulation as AI-CPS}
As shown in Fig.~\ref{fig:CPS_model}, a CPS typically comprises four major components: a physical plant $M$, a software control unit, an actuator, and a network of sensors.
Currently, AI technologies have been applied across various components in CPSs, such as perception units and decision-making units.
In this work, we specifically focus on AI-CPSs that use DRL algorithms as AI software controllers to achieve real-time control of a physical plant $M$.
The control command $u$ is computed by the AI controller $C$, which makes control decisions based on the current state $y$ and an external input $i$.
After receiving the control command $u$, the plant evolves to the next state according to the dynamics of the physical system.

As a typical example of CPS, the robotic manipulator integrates physical components, e.g., the robot arm and the sensors, with digital devices, e.g., the software controller, to perform tasks in the real world.
The dynamics of the manipulator is often described in joint space as
\begin{equation}
    M(\boldsymbol{q})\boldsymbol{\dot{q}} + C(\boldsymbol{q}, \boldsymbol{\dot{q}}) + G(\boldsymbol{q}) = \boldsymbol{\tau}   
\end{equation}
where $\boldsymbol{q} = [q_1,q_2,\ldots,q_n]^T$ and $\boldsymbol{\dot{q}} = [\dot{q}_1,\dot{q}_2,\ldots,\dot{q}_n]^T$ are the vectors of joint angles and angular velocities, respectively.
$n$ indicates the degrees of freedom (DoF) of the robotic manipulator.
$M(\boldsymbol{q})$, $C(\boldsymbol{q}, \boldsymbol{\dot{q}})$ and $G(\boldsymbol{q})$ are the mass matrix, the vector of centrifugal and Coriolis terms, and the gravity terms, respectively.
$\boldsymbol{\tau}$ represents the torques applied on each joint of the manipulator. 

By taking the current system states, e.g., $\boldsymbol{q}$ and $\boldsymbol{\dot{q}}$, as well as external sensor information, e.g., the status of the manipulation objects, as the input, the control objective of robotics manipulation is thus to design a controller $C$ that generates a sequence of control commands $u = \boldsymbol{\tau}$ to fulfill the desired task requirements, 
However, traditional software controllers usually rely on accurate modelling of the system behavior, which in complicated tasks, e.g. manipulating deformable objects, is hard to achieve.
Moreover, designing a controller based on a specific model limits its capability of performing well in different environments and tasks, which further diminishes its applicability and generalizability. 

To address these challenges, recent studies in robotics manipulation leverages AI techniques, e.g., DRL~\cite{arulkumaran2017deep}, to learn a control policy directly from data, making it a typical and representative example of AI-CPS. 
Therefore, by focusing on evaluating and testing these industrial-level AI-enabled robotics manipulation tasks, we are able to also obtain a thorough understanding of how modern AI-CPS will perform in actual critical industrial domains.  


\subsection{STL and Falsification}

STL~\cite{donze2010robust} is a specification language designed to describe the expected \emph{temporal}-related behavior of a system, e.g., safety and performance. 
It is equipped with \emph{quantitative robust semantics}, which quantitatively measures the degree of satisfaction of the specification.
Formally, given an STL specification $\varphi$ and a system output $M(i)$, where $i$ is the input signal, the STL semantics $\textsc{rob}(M(i), \varphi)$ maps $M(i)$ and $\varphi$ to a real number. 
A positive/negative value of $\textsc{rob}(M(i), \varphi)$ represents that the specification is satisfied/violated, and a larger value implies a stronger satisfaction/violation.
Interested readers are referred to the work of Donze et al.~\cite{donze2010breach} for a detailed description of the robust semantics.

As a widely used safety assurance technique, \emph{falsification}~\cite{zhang2018two,zhang2022falsifai,deshmukh2017testing,donze2010breach,zhang2021effective} aims to discover input signals to a system that could violate a desired specification.
It transforms the discovery process into an optimization problem, with the goal of minimizing the robust semantics such that $\textsc{rob}(M(i), \varphi)$ is less than zero. 
Hill-climbing based optimization methods~\cite{yuret1993dynamic}, e.g., dual annealing~\cite{xiang1997generalized} and Nelder-Mead~\cite{gao2012implementing}, are commonly used to solve this problem.
Typically, the optimization process involves first providing some initial samples to the system and then proposing new samples based on the feedback received, which is represented by the robust semantics in this case.

\section{Research Questions}

Considering the scope of this paper, we investigate the following three research questions (RQs) to have a better understanding of the challenges and opportunities of building reliable AI-enabled industrial-level robotics manipulation systems utilizing Isaac Sim.


\noindent \textbf{RQ1: What are the advantages and limitations of Isaac Sim compared to other physical simulators?} 

Although Isaac Sim is recognized for its seamless integration with NVIDIA hardware, the requirements and challenges in simulating robotics manipulations are diverse and often dependent on the tasks. 
In the field of robotics, there are numerous physical simulators, such as Gazebo~\cite{1389727}, PyBullet~\cite{benelot2018}, and Mujoco~\cite{todorov2012mujoco}, that have been widely used by researchers and practitioners to meet their specific demands. 
As a novel simulation platform that is still in its developmental stages, there is a need for an analysis of the strengths and limitations of Isaac Sim compared to other physical simulators, as well as its performance in simulating various robotics tasks.

Therefore, to study RQ1, we conducted a survey with industrial and academic practitioners around the world to gather their comments on the advantages and limitations of Isaac Sim as compared to other physical simulators. 
The purpose of the survey is to gain insights into the unique features and benefits of Isaac Sim that may make it stand out from other simulators, as well as to identify any potential drawbacks or areas where improvements could be made.
Details about the RQ1 are presented in Sec.~\ref{sec.pros_cons_isaacsim}.

\noindent \textbf{RQ2: How do AI controllers perform in various robotics manipulation tasks?}

The findings from RQ1 (see Sec.~\ref{subsec:survey_results} for details) highlight the necessity of an industrial-level benchmark that aids users in initiating and familiarizing themselves with the utilization of Isaac Sim for the development of AI-enabled robotics applications.
To address this requirement, with our industrial partners, we develop a benchmark comprising eight typical robotics manipulation tasks that fully cover rigid body manipulation, deformable object manipulation, prehensile manipulation, and non-prehensile manipulation, using Isaac Sim. 

In the meantime, despite extensive research on DRL controllers in robotics, there is still a lack of systematic and detailed analysis of their performance in diverse robotics manipulation tasks. 
This knowledge gap emphasizes the need to compare and evaluate different DRL controllers under various task requirements. 
Therefore, we design RQ2 that aims to compare different DRL controllers across a wide range of manipulation tasks. 
For this purpose, we propose various metrics to evaluate and examine the performance of different DRL controllers on the tasks presented in our benchmark.
The result reveals the strengths and limitations of different AI controllers in robotics manipulation and allows researchers to identify the most effective methods for achieving specific control objectives. 
Details about this RQ are presented in Sec.~\ref{sec.benchmark}.

\noindent \textbf{RQ3: How effective are different optimization methods in falsifying physics engine-based robotics manipulation tasks?}

The findings of RQ1 also emphasize the need for testing support in the development of AI-enabled robotics applications with Isaac Sim.
Building upon this insight, we develop a Python-based falsification framework that can be directly used with physical simulators as well as OpenAI Gym environments.
However, although falsification has proved to be effective on traditional CPSs, its efficacy in the context of robotics tasks simulated using modern physical simulators remains unclear.
To address this gap, we introduce RQ3 to analyze the performance of different optimization methods, i.e., random, Nelder-Mead, and dual annealing, in falsifying robotics tasks with AI software controllers. 
By using our benchmark, we conduct a falsification test that reveals the robustness and reliability of AI controllers and helps identify potential failures and vulnerabilities in the system. 
Additionally, this test also demonstrates the extensibility and applicability of our benchmark.
Details about RQ3 are given in Sec.~\ref{sec.falsification_tool}.

\noindent{\textbf{Hardware \& Software Dependencies.}}
The simulation, DRL training and falsification were conducted using SKRL~\cite{serrano2022skrl}, RTAMT~\cite{nivckovic2020rtamt} and Scipy~\cite{2020SciPy-NMeth} libraries.
For the hardware platform, we use three Lambda Tensorbooks, each of which has an Intel(R) Core(TM) i7-10870H CPU @ 2.20GHz Processor with 8 CPUs and an NVIDIA RTX 3080 Max-Q GPU with 16 GB VRAM. 
The total computation time of training and testing takes over 600 hours.

\section{Advantages and Limitations of Isaac Sim}
\label{sec.pros_cons_isaacsim}

As a first step towards understanding the demands of industrial and academic practitioners, we conduct a survey to get opinions about the advantages and limitations of Isaac Sim. 
In this section, we describe the survey setting (Sec.~\ref{subsec:survey_setting}) and present the results (Sec.~\ref{subsec:survey_results}).

\subsection{Survey Setting}
\label{subsec:survey_setting}

\noindent \textbf{Survey Design.}
The survey consists of three question types: short-answer, multiple-choice, and Likert scale questions. 
The Likert scale questions ask respondents to indicate their level of agreement with statements such as Strongly Agree, Agree, Neutral, Disagree, Strongly Disagree, and I don't know.
The survey consists of five parts:
\begin{compactitem}[$\bullet$]
    \item \textit{Demographics:} We aim to gather information about the practitioners' demographics and experience in robotics development.

    \item \textit{General robotics simulation using Isaac Sim:} 
    We target to investigate the pros and cons of Isaac Sim in terms of general robotics simulation and DRL-related modules.
    
    \item \textit{Development challenge of DRL robotics tasks in Isaac Sim:}
    In this part, we ask about the practitioners' experience in developing DRL robotics tasks in Isaac Sim.
    
    \item \textit{Comparison with other robotics simulators:}
    This part asks the practitioners to compare Isaac Sim with other popular robotics simulators, such as Gazebo and PyBullet.
    
    \item \textit{Testing of robotics tasks in Isaac Sim:}
    This part studies the practitioners' habits in performing testing in robotics tasks and the status of testing support in Isaac Sim.
\end{compactitem}
We also provide an open-ended question at the end of the survey, inviting practitioners to share their opinions and comments about Isaac Sim in the context of robotics and DRL development.

\noindent \textbf{Participant Recruitment.}
We recruited practitioners through two methods: email contact and forum recruitment.
To obtain email contacts, we searched for research papers published in top conferences such as ICSE, ESEC/FSE, ASE, ICRA, IROS, AAAI, IJCAI, and NeurIPS from 2019 to 2023 that involve developing robotics tasks in Isaac Sim. 
We collected the email addresses of the authors of these papers, resulting in a list of 75 researchers whom we contacted individually via email.
For forum recruitment, we posted topics about the survey in various online communities, including the Isaac Sim forum~\cite{isaacforum}, the NVIDIA official Discord community, and the GitHub discussion page of relevant libraries, e.g., SKRL~\cite{serrano2022skrl} and Isaac Gym~\cite{makoviychuk2021isaac}.
In total, we received 17 responses from nine different countries. 
More details about the information of the respondents can be found on our website\footnote{\website}.



\subsection{Result Analysis and Findings}
\label{subsec:survey_results}

\begin{figure*}
\centering
\includegraphics[width=\linewidth]{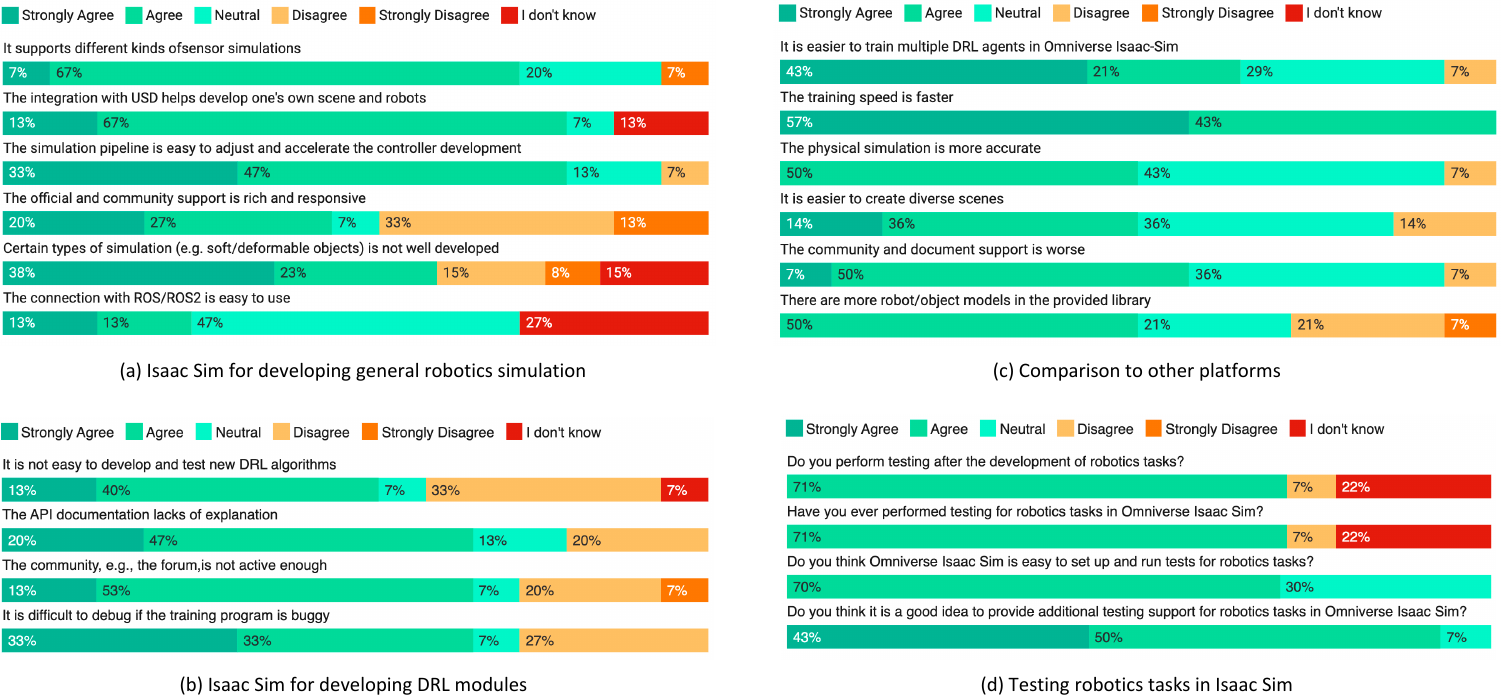}
\caption{RQ1 - Partial questions and results from survey}
\vspace{-10pt}
\label{fig:RQ1_result}
\end{figure*}

Fig.~\ref{fig:RQ1_result} presents the survey results from four important perspectives (the complete version can be found on our website).
As depicted in Fig.~\ref{fig:RQ1_result}(a), the majority of practitioners agree that Isaac Sim supports diverse types of sensors (74\%), and the USD (Universal Scene Design) file format~\cite{usd} makes it easier to develop custom scenes and robots (80\%).
However, some respondents point out that certain simulation features, such as simulating soft or deformable objects, are not well developed.

Regarding challenges in developing DRL tasks in Isaac Sim, Fig.~\ref{fig:RQ1_result}(b) shows that most practitioners think that the API documentation lacks proper explanation (67\%) and the community support is not active enough (66\%), which highlights the need for improved documentation and official support for users of Isaac Sim.
Moreover, more than half (53\%) of the practitioners consider that it is not easy to develop and test new DRL algorithms. 

In terms of comparison with other platforms (see Fig.~\ref{fig:RQ1_result}(c)), all practitioners believe that the training speed in Isaac Sim is faster than other simulation platforms due to its good compatibility with NVIDIA GPUs. 
Additionally, most of them (63\%) find it easier to train multiple DRL agents in Isaac Sim. 
However, more than half of the practitioners (57\%) think that the community and documentation support is not as good as in other platforms. 
Moreover, as the simulation prototype is ultimately expected to be deployed in the real world, we are also interested in the accuracy of Isaac Sim's physical simulation.
While half of the practitioners think that the simulation is more accurate, the other half are neutral or disagree with this point. 

In addition, we ask practitioners about their opinions on testing robotics tasks in Isaac Sim (see Fig.~\ref{fig:RQ1_result}(d)).
Most of them (93\%) believe that it is a good idea to provide additional testing support, with one practitioner stating, "\textit{An advanced plug-and-play testing module will be an asset.}".
This encourages our development of a falsification framework for Isaac Sim and OpenAI Gym environments.

Conclusively, numerous participants have pointed out the lack of a clear pipeline for DRL and robotics development in the current version of Isaac Sim.
Notably, many responses indicate that users face challenges in initiating their development process due to insufficient instructions and documentation. 
Moreover, the absence of a baseline for comparison and calibration poses difficulties for developers who lack guidance and standards to assess their progress accurately.
Therefore, we consider an AI-enabled robotics manipulation benchmark can greatly benefit practitioners in the following aspects: 
1) an easy-to-use development playground and pipeline;
2) a standard baseline for observation and comparison; 
and 3) an extensible and applicable environment for quick demonstration and prototyping.




\begin{tcolorbox}[size=title, colback=white]
{\textbf{Answer to RQ1:} 
Isaac Sim has gained recognition for its faster training speed and support for various sensors and custom scenes. 
However, practitioners have expressed concerns regarding the lack of a clear pipeline for DRL training and robotics development. 
A benchmark of AI-enabled robotics manipulation is desired.
Additionally, there is a demand for testing support within Issac Sim. 
}
\end{tcolorbox}

\section{Benchmark of Robotics Manipulation}
\label{sec.benchmark}


As revealed by Sec.~\ref{subsec:survey_results}, practitioners are seeking a versatile and user-friendly environment that facilitates quick prototyping and demonstration of industrial AI-enabled robotics systems. 
Consequently, we develop an industrial-level public benchmark for AI-enabled robotics manipulation with high extensibility and applicability as a timely solution to this demand. 
In this section, we first describe how the benchmark is construed based on Isaac Sim (Sec.~\ref{subsec:benchmark_construction}).
Then, by using the developed benchmark, we systematically evaluate the performance of AI software controllers.
In particular, we introduce four evaluation metrics in Sec.~\ref{subsec:eval_metrics} and present the evaluation results in Sec.~\ref{sec.benchmark_result}.

\subsection{Benchmark Construction}
\label{subsec:benchmark_construction}

\begin{figure*}[!t]
\centering
\subfloat[Point Reaching (PR)]{\includegraphics[width=1.8in, height = 1.8in]{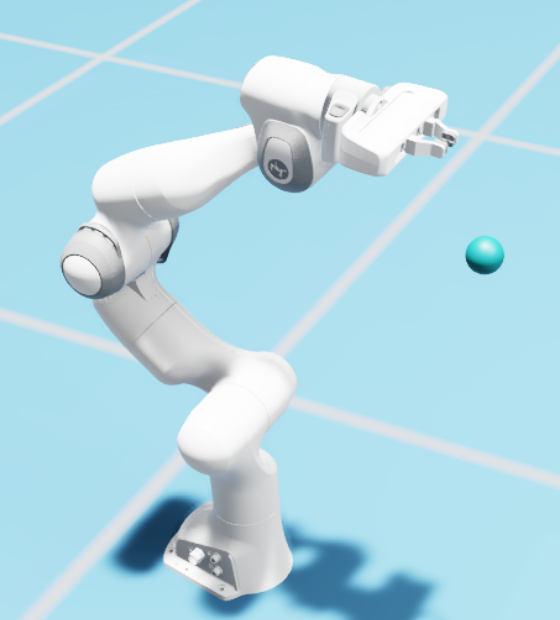}%
\label{fig_PR}}
\hfil
\subfloat[Cube Stacking (CS)]{\includegraphics[width=1.8in, height = 1.8in]{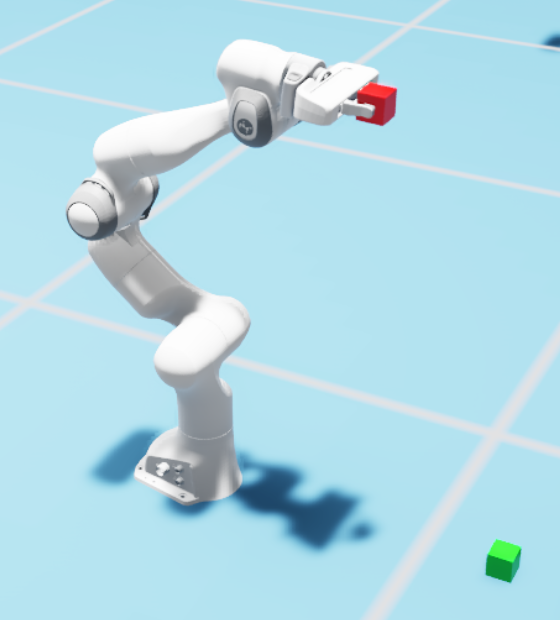}%
\label{fig_CS}}
\hfil
\subfloat[Peg-in-Hole (PH)]{\includegraphics[width=1.8in, height = 1.8in]{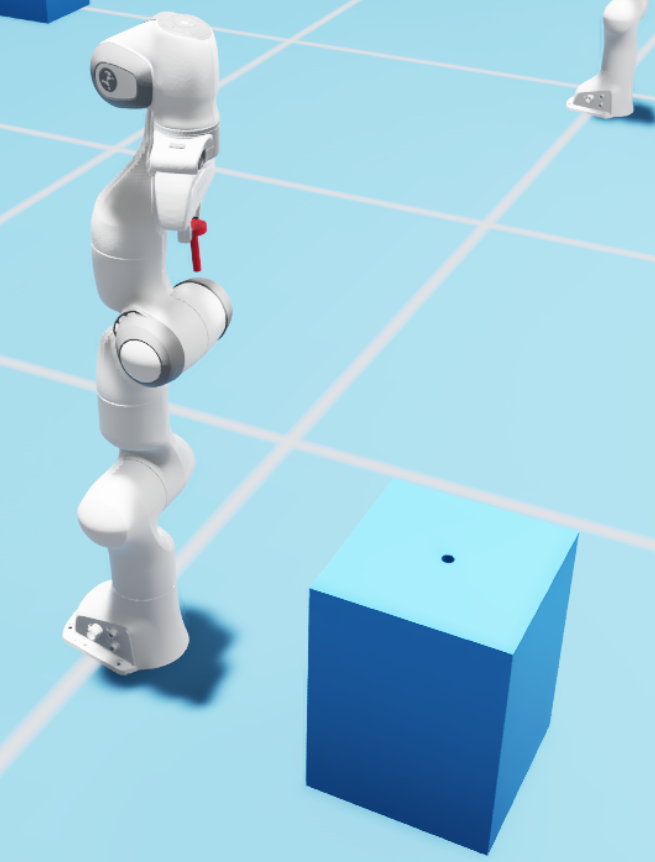}%
\label{fig_PH}}
\hfil \\
\subfloat[Ball Balancing (BB)]{\includegraphics[width=1.8in, height = 1.8in]{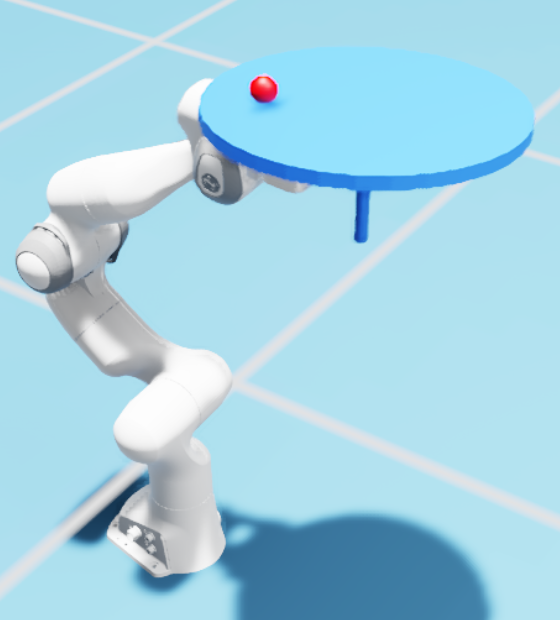}%
\label{fig_BB}}
\hfil 
\subfloat[Ball Catching (BC)]{\includegraphics[width=1.8in, height = 1.8in]{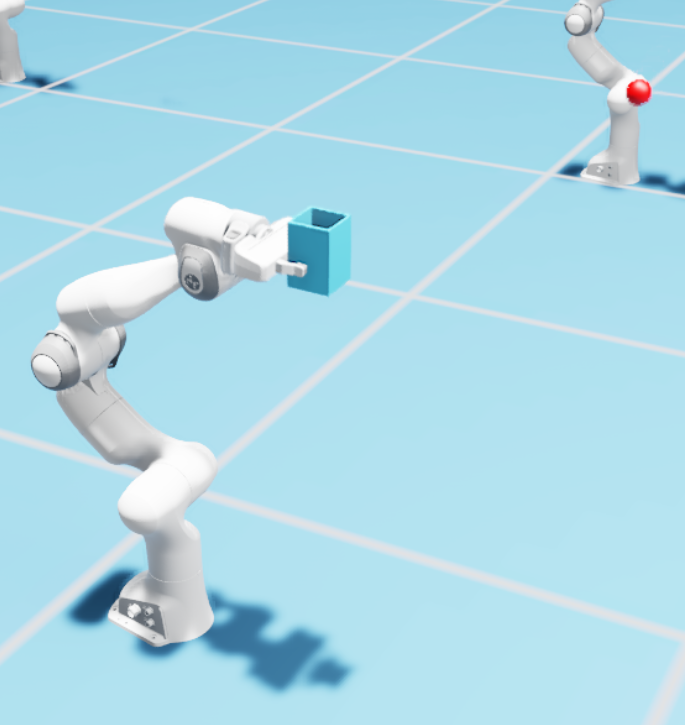}%
\label{fig_BC}}
\hfil
\subfloat[Ball Pushing (BP)]{\includegraphics[width=1.8in, height = 1.8in]{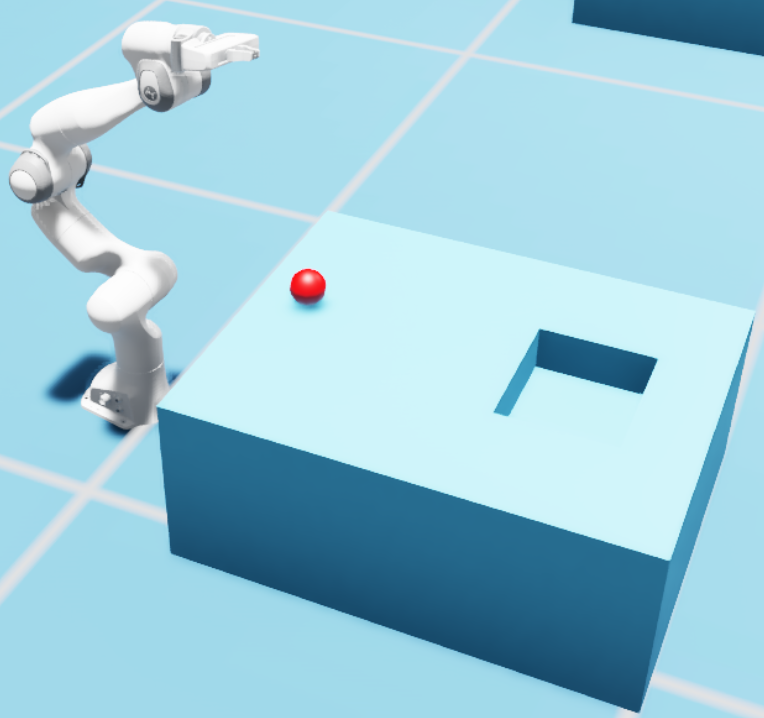}%
\label{fig_BP}}
\hfil \\
\subfloat[Door Openning (DO)]{\includegraphics[width=1.8in, height = 1.8in]{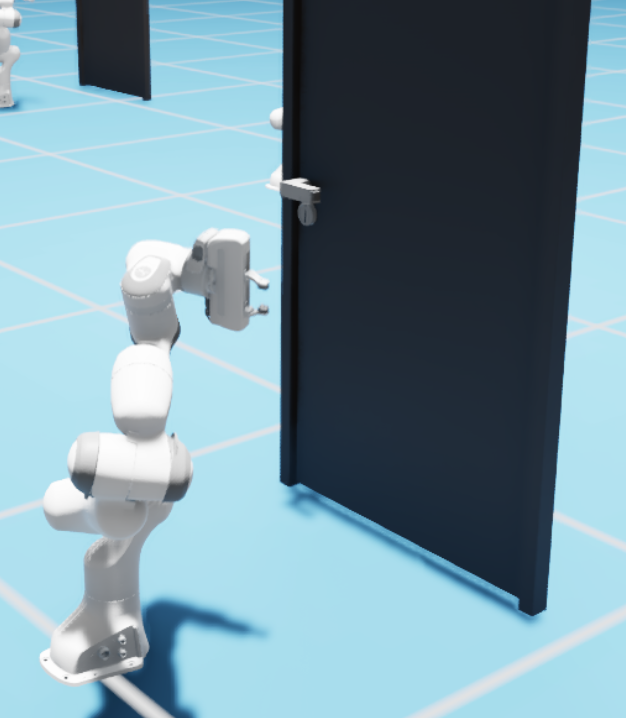}%
\label{fig_DO}}
\hfil
\subfloat[Cloth Placing (CP)]{\includegraphics[width=1.8in, height = 1.8in]{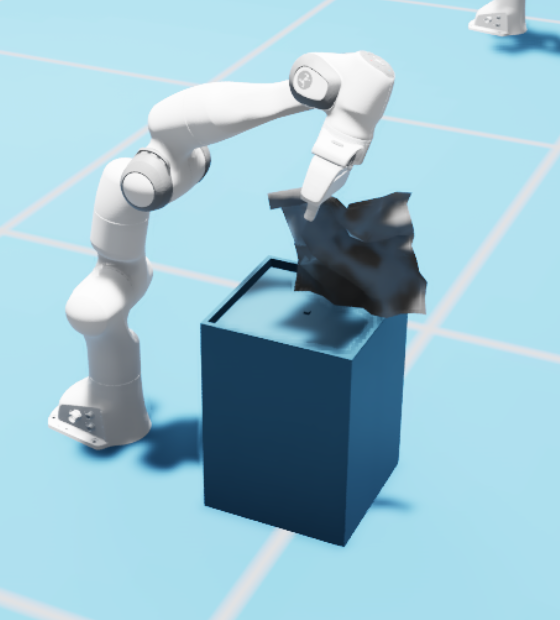}%
\label{fig_CP}}
\caption{Robotics manipulation tasks included in the proposed benchmark based on Isaac Sim.}
\label{fig_benchmark_tasks}
\end{figure*}


Our benchmark includes eight typical and diverse industrial-level robotics manipulation tasks (see Fig.~\ref{fig_benchmark_tasks}). 
The manipulator model used in all tasks is the Franka Emika Panda~\cite{franka}, a state-of-the-art robotic arm widely used for various manipulation tasks in research and industry~\cite{gaz2019dynamic,petrea2021interaction,zhang2020modular}. 
It is equipped with seven DoFs, a high-resolution torque sensor in each joint, and an advanced control system that allows for precise and smooth motion. 
The selected tasks cover a broad range of scenarios, including rigid body manipulation, soft/deformable object manipulation, non-prehensile manipulation, etc. 
Each task requires a unique set of control strategies and presents a varying level of difficulty, ranging from simple to complex, to test the performance and adaptivity of AI controllers. 
A detailed introduction to the presented benchmark is given as follows.

\noindent \textbf{Tasks.} The benchmark contains a total of eight tasks, each designed to test different aspects of a robot's manipulation capabilities. 

\begin{compactitem}[$\bullet$]
    \item \textit{Point Reaching (PR):} The robot needs to reach a specific point in 3D Cartesian space by using its end-effector. 
    This task is a fundamental and essential functionality that a manipulator should possess. 
    
    \item \textit{Cube Stacking (CS):} In this task, the robot needs to grasp a cube-shaped object and accurately place it on top of another cube-shaped object that serves as the target. 
    This pick-and-place like task is a common task in industrial settings, e.g., assembly lines.

    \item \textit{Peg-in-Hole (PH):} The robot needs to accurately insert a cylindrical object into a corresponding hole. 
    This task is commonly used in manufacturing processes, such as circuit board assembly.

    \item \textit{Ball Balancing (BB):} The objective of this task is to balance a ball at the center of a tray that is held by the robot's end-effector.
    This task is useful for applications such as stabilizing a moving platform.

    \item \textit{Ball Catching (BC):} The robot needs to catch a ball that is thrown to it using a tool. 
    This task demands the capability to handle moving objects. 

    \item \textit{Ball Pushing (BP):} The robot is required to push a ball towards a target hole on a table. 
    As a typical non-prehensile manipulation, this task is important in many industrial applications, such as material handling and conveyor systems.

    \item \textit{Door Opening (DO):} The robot attempts to open a door using its gripper. 
    This is a more advanced manipulation task and requires a multi-stage control process. 
    
    \item \textit{Cloth Placing (CP):} The robot has to move and place a piece of cloth onto a target table. 
    As a soft object manipulation task, it often requires a more complex controller than rigid body object manipulations.
    
\end{compactitem}



\noindent \textbf{AI software controllers.} We employ various DRL algorithms, such as \emph{Trust Region Policy Optimization} (TRPO)~\cite{schulman2015trust}, \emph{Deep Deterministic Policy Gradient} (DDPG)~\cite{lillicrap2015continuous}, \emph{Soft Actor-Critic} (SAC)~\cite{haarnoja2018soft}, \emph{Proximal Policy Optimization} (PPO)~\cite{schulman2017proximal}, and \emph{Twin-Delayed Deep Deterministic} (TD3)~\cite{fujimoto2018addressing}, as the AI software controller of the robotic manipulator. 
These algorithms are based on the implementation given in the SKRL library~\cite{serrano2022skrl}. 
The AI controller takes the states of the manipulator and the object to be manipulated as input and generates the control command for each robot joint. 
For each task, we design a specific reward function for training the AI controllers, which is detailed on our website.

\noindent \textbf{Learning Environments.} To facilitate training and evaluation, we wrap all tasks within the Omniverse Isaac Gym Reinforcement Learning Environment~\cite{oige}, which is built on top of the OpenAI Gym framework~\cite{1606.01540}. 
This provides better compatibility with Isaac Sim simulations and other DRL libraries and allows for easy extensions in the future.

\noindent \textbf{Initial Configurations.} By treating each manipulation task as an independent entity, we consider the initial configuration of either the object to be manipulated or the target object as the input signal to the system.
During the training process of the AI controller, we vary the initial configuration to evaluate and test its performance.
The allowable range of initial configurations for each task used in this paper is specified in Table~\ref{table_performance}.
All position values represent the relative Cartesian distances to the base of the manipulator.

\subsection{Evaluation Metrics}
\label{subsec:eval_metrics}

\begin{table*}[t]
\centering
\caption{Performance evaluation of AI controllers in robotics manipulation tasks (better results are highlighted in gray).}
\label{table_performance}
\resizebox{1\textwidth}{!}{
\begin{tabular}{c|c|c|r|rrr|rrr}
\toprule
& & & & \multicolumn{3}{c|}{No noise} & \multicolumn{3}{c}{With noise}      \\
\textbf{Task} &  \textbf{Initial Configuration} &\textbf{Controller}& \textbf{TT} & \textbf{SR} & \textbf{DBR} & \textbf{TCT} & \textbf{SR} & \textbf{DBR} & \textbf{TCT}\\ \midrule
\multirow{2}{*}{PR} & Position of the target point ($x,y,z$-coordinates) & PPO &  \tbmgray 10500 & 93\% & 0.26\% & 29.29 & 87\% & 0.27\% &  29.83  \\
                 & $x\in [0.3, 0.7]$ $y\in [-0.4,0.4]$, $z\in [0.4,0.8]$   & TRPO & 26700 & \tbmgray 100\% & \tbmgray 0.07\% & \tbmgray 16.21 & \tbmgray 98\% & \tbmgray 0.12\% & \tbmgray 18.39  \\  \hline
\multirow{2}{*}{CS} & Position of the target cube ($x,y$-coordinates)  & PPO & \tbmgray 11300 & \tbmgray 100\% & \tbmgray 7.13\% & 75.66 & 98\% & 24.77\% & 75.74 \\
                  & $x\in [0.4, 0.8]$, $y\in [-0.1, 0.3]$  & TRPO & 12600 & 99\% & 8.98\% & \tbmgray 66.44 & 98\% & \tbmgray 15.62\% & \tbmgray 66.80   \\ \hline
\multirow{2}{*}{PH} & Position of the target hole ($x,y$-coordinates)  & PPO &  \tbmgray 154000 & 85\% & 3.59\% & \tbmgray 16.79 & 78\% & 4.41\% & \tbmgray 17.83\\
                   & $x\in [0.3, 0.7]$, $y\in [-0.2, 0.2]$ & TRPO & 259600 & \tbmgray 92\% & \tbmgray 1.90\% & 43.53 & \tbmgray 90\% & \tbmgray 2.41\% & 46.57\\ \hline
\multirow{2}{*}{BB} & Initial position of the ball ($x,y$-coordinates)   & PPO & \tbmgray 11500 & 98\% & 0.47\% & 6.41 & 89\% & 3.44\% & 10.24\\
                   & $x\in [0.2, 0.5]$, $y\in [-0.15, 0.15]$ & TRPO & 12900 & \tbmgray 100\% & \tbmgray 0.01\% & \tbmgray 5.91 & \tbmgray 93\% & \tbmgray 2.23\% & \tbmgray 6.01\\ \hline
\multirow{2}{*}{BC} & Initial position of the ball ($x,y$-coordinates)  & PPO &  \tbmgray 20800 & \tbmgray 100\% & \tbmgray 6.34\% & 23.14 & \tbmgray 99\% & \tbmgray 6.96\% & 23.39\\
                   & $x\in [1.05, 1.15]$, $y\in [-0.05, 0.05]$ & TRPO & 28100 & 97\% & 8.69\% & \tbmgray 21.60 & 93\% & 9.14\% & \tbmgray 21.86\\ \hline
\multirow{2}{*}{BP} & Initial position of the ball ($x,y$-coordinates)  & PPO &  \tbmgray 60300 & 98\% & 9.28\% & 53.53 & 97\% & 13.78\% & 55.45\\
                    & $x\in [0.4, 0.6]$, $y\in [-0.1, 0.1]$ & TRPO & 175100 &\tbmgray 100\% & \tbmgray 6.85\% & \tbmgray 37.12 & \tbmgray 100\% & \tbmgray 7.88\% & \tbmgray 38.27\\ \hline
\multirow{2}{*}{DO} & Position of the door ($x,y$-coordinates)  & PPO &  \tbmgray 124400 & 89\% & 36.60\% & 136.62 & 81\% & 39.22\%  & 154.63\\
                  & $x\in [0.75, 0.85]$, $y\in [-0.1, 0.1]$  & TRPO & 128300 & \tbmgray 97\% & \tbmgray 31.11\% &\tbmgray 126.36 & \tbmgray 94\% & \tbmgray 32.74\% & \tbmgray 127.06\\ \hline 
\multirow{2}{*}{CP} & Position of the target table ($x,y$-coordinates)  & PPO &  34700 & \tbmgray 100\% & \tbmgray 2.44\% & \tbmgray 12.15 & 99\% & \tbmgray 2.85\% & \tbmgray 14.87\\
                  & $x\in [0.45, 0.75]$, $y\in [-0.35, 0.35]$   & TRPO & \tbmgray 22400 & 99\% & 14.38\% & 27.45 & 99\% & 15.08\% & 29.25\\ 
                    \bottomrule
\end{tabular}
}
\end{table*}

With the developed benchmark of robotics manipulation tasks, we are able to answer RQ2.
For this purpose, we perform the evaluation based on four categories of evaluation metrics that consider different aspects of the manipulation tasks:

\begin{compactitem}[$\bullet$]
    \item \textit{Success Rate (SR):} This metric measures the percentage of successful task completions among all attempted trials. 
    Successful task completion is defined as when the controlled system behavior satisfies a predefined task-relevant STL specification (see Table~\ref{table_task_STL} for the STL specifications used in our benchmark).
    A high SR indicates that the AI controller is capable of effectively completing the task. 

    \item \textit{Dangerous Behavior Rate (DBR):} DBR is computed as the percentage of time steps, i.e., control intervals, where the manipulator is close to failing the task, among all simulated time steps.
    A high DBR indicates that the AI controller is prone to generating unsafe or unstable control commands, which can potentially cause task failures.
    The STL specifications used in our benchmark for describing dangerous behaviors are given in Table~\ref{table_task_STL}.

    \item \textit{Task Completion Time (TCT):} This metric measures the time steps needed by the manipulator to successfully complete the task. 
    A shorter TCT indicates that the AI controller is capable of generating more efficient control commands to complete the task.

    \item \textit{Training Time (TT):} We use TT to measure the training time steps required to train the AI controller, i.e., the policy reward converges to a constant level. 
    A shorter training time indicates that the used DRL algorithm is more efficient in learning the control policy for the given task.
    
\end{compactitem}


Note that, in real-world applications, system noises and uncertainties, such as sensor noises or model inaccuracies, are often inevitable.
Thus, the ability of manipulators to remain robust under such conditions is crucial, especially for those designed for real-world tasks.
Therefore, we include \textit{robustness} as our final critical evaluation metric.
Specifically, to assess the robustness of AI controllers, we introduce action noises to the trained AI controllers and measure how the SR, DBR, and TCT are impacted.
The results of this analysis are presented in the following subsection.

\subsection{Experimental Evaluations}
\label{sec.benchmark_result}

\begin{figure}
  \centering
  \includegraphics[width=.8\linewidth]{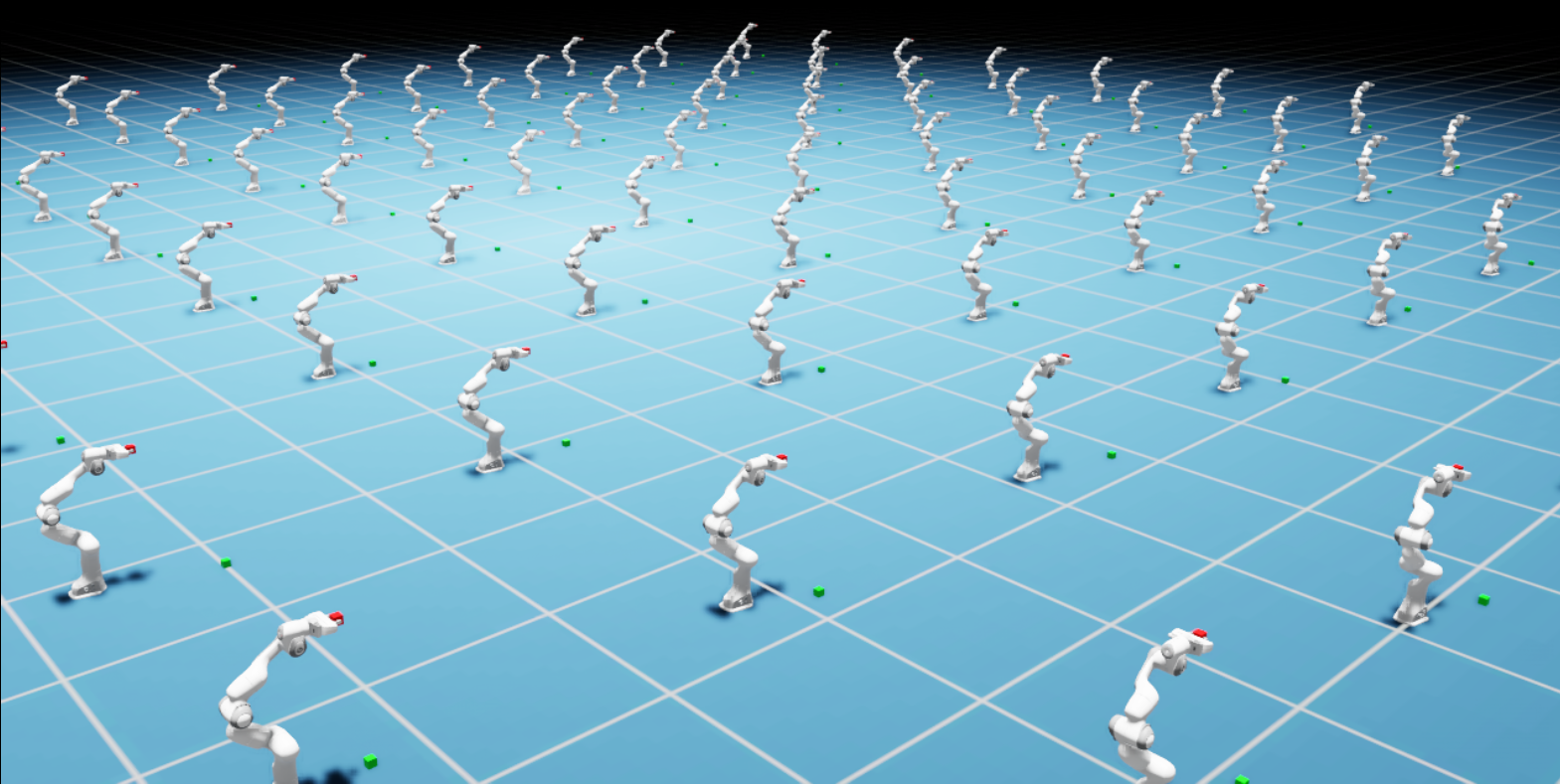}
  \caption{Parallel running of environments in Isaac Sim.}
  \label{fig_parallel}
\end{figure}

\begin{table*}[t]
\centering
\caption{STL specifications for different robotics manipulation tasks ($\Box$: Globally. $\Diamond$: Eventually).}
\label{table_task_STL}
\resizebox{1\textwidth}{!}{
\setlength{\tabcolsep}{10pt}
\begin{tabular}{ccc}
\toprule
\textbf{Task} & \textbf{Successful Task Completion} & \textbf{Dangerous Behavior}  \\ \hline
PR & $\Box_{[0, 30]}(\lVert\mathtt{finger_{pos}} -\mathtt{point_{pos}}\rVert \le 0.3)$ & $\Box_{[0, 30]}(\lVert\mathtt{finger_{pos}} -\mathtt{point_{pos}}\rVert \le 0.12)$ \\ \hline
\multirow{3}{*}{CS} & 
\multirow{3}{*}{\shortstack{$\Diamond_{[0, 30]}(\lVert\mathtt{cube_{pos}} -\mathtt{target_{pos}}\rVert \le 0.024 \wedge$ \\ $ \lVert\mathtt{cube_{pos\_z}} -\mathtt{target_{pos\_z}}\rVert \ge 0 ) $} } 
& \multirow{3}{*}{\shortstack{$\Box_{[0, 30]}(\lVert\mathtt{cube_{pos}} -\mathtt{target_{pos}}\rVert \le 0.35 \vee$ \\ $ \lVert\mathtt{cube_{pos\_z}} -\mathtt{target_{pos\_z}}\rVert \ge 0.02) $} }  \\ \\ \\ \hline
PH & $\Box_{[25, 30]}(\lVert\mathtt{obj_{pos}} -\mathtt{hole_{pos}}\rVert \le 0.12)$ & $\Box_{[25, 30]}(\lVert\mathtt{obj_{pos}} -\mathtt{hole_{pos}}\rVert \le 0.37)$ \\ \hline
BB & $\Box_{[5, 20]}(\lVert\mathtt{ball_{pos}} -\mathtt{tray_{pos}}\rVert \le 0.25)$ & $\Box_{[5, 20]}(\lVert\mathtt{ball_{pos}} -\mathtt{tray_{pos}}\rVert \le 0.2)$ \\ \hline
BC & $\Box_{[5, 30]}(\lVert\mathtt{ball_{pos}} -\mathtt{tool_{pos}}\rVert \le 0.1)$ &  $\Box_{[5, 30]}(\lVert\mathtt{ball_{pos}} -\mathtt{tool_{pos}}\rVert \le 0.2)$ \\ \hline
BP & $\Diamond_{[0, 30]}(\lVert\mathtt{ball_{pos}} -\mathtt{hole_{pos}}\rVert \le 0.3)$ & $\Diamond_{[0, 30]}(\lVert\mathtt{ball_{pos}} -\mathtt{hole_{pos}}\rVert \le 0.5)$ \\ \hline
DO & $\Diamond_{[0, 30]}(\mathtt{door_{yaw}} \ge 20)$ & $\Diamond_{[0, 30]}(\mathtt{door_{yaw}} \ge 0.1)$ \\ \hline
\multirow{3}{*}{CP} & \multirow{3}{*}{$\Diamond_{[0, 30]}(\lVert\mathtt{cloth_{pos}} -\mathtt{table_{pos}}\rVert \le 0.25)$} 
& \multirow{3}{*}{\shortstack{$\Box_{[0, 30]}(\lVert\mathtt{cloth_{pos}} -\mathtt{table_{pos}}\rVert \le 0.3 \vee$ \\ $\lVert\mathtt{cloth_{pos\_z}} -\mathtt{ground_{pos\_z}}\rVert \ge 0.02)$}} \\ \\ \\ \bottomrule
\end{tabular}
}
\end{table*}



\noindent \textbf{Experiment Settings.} 
To evaluate the performance of AI-enabled robotics manipulation, we first use various DRL algorithms, i.e., TRPO, PPO, SAC, TD3, and DDPG, to train multiple AI software controllers for each manipulation task. 
However, our experiments reveal that except for TRPO and PPO, the other DRL algorithms fail to produce a working controller that is capable of solving the manipulation task. 
One potential explanation for this could be that, compared to off-policy algorithms, on-policy algorithms are better suited for Isaac Sim, which employs parallel running of a large number of environments (see Fig.~\ref{fig_parallel}). 
In such a case, the collected samples are strongly correlated in time, and a considerable amount of information is accumulated in each time step, posing challenges for off-policy algorithms. 
Further research is needed to investigate whether this is the root cause of the issue. 
Consequently, we use only the AI controllers trained by PPO and TRPO in our performance evaluation and the falsification test presented in the next section, as they are the only functional ones.

For each task and AI controller, we conduct 100 trials separately under two different conditions: without and with action noise, where the action noise is a white Gaussian noise with a variance of 0.25.
For each trial, the initial configuration is generated randomly according to Table~\ref{table_performance}, and the simulation length is set to 300 time steps.
The values of DBR and TCT are averaged among all trails.

\noindent \textbf{Experiment Results.} 
The results are presented in Table~\ref{table_performance}.
It can be observed that both TRPO and PPO are able to accomplish most of the manipulation tasks with a SR over $90\%$, except for PPO in PH ($85\%$) and DO ($89\%$). 
While TRPO maintains a comparable SR to PPO in tasks like CS, BC, and CP, it has a better performance in other tasks, particularly for those that require precise control (e.g., PH) or a multi-stage control process (e.g., DO). 
As a result, TRPO generally outperforms PPO in terms of SR.
Similar trends can also be observed in the metrics DBR and TCT, where TRPO usually has a better performance with the exception that for the task CP, PPO has a more clear advantage.
However, TRPO often requires a longer TT, especially for tasks PR, PH, and BP, where as much as twice of the TT is required compared to PPO. 

As expected, the introduction of action noise leads to a decrease in the performance of AI controllers across all metrics. 
However, both PPO and TRPO are still able to accomplish the tasks with a satisfactory SR, where the decrease is less than 5\% for most of the tasks. 
Tasks that require accurate execution of actions, such as PH, BB, and DO, exhibit a more noticeable decrease in SR. 
In terms of the DBR, the action noise has a strong impact on the CS and BB tasks. 
One potential reason could be that these tasks have less tolerable space for manipulating objects, e.g., the size of the target cube in CS or the tool in BB, making imprecise actions more dangerous. 
The impact of the action noise on the DBR of other tasks as well as on the TCT is marginal. 
Overall, both PPO and TRPO show good performance in terms of robustness against action noise.


\begin{tcolorbox}[size=title, colback=white]
{\textbf{Answer to RQ2:} The AI controllers trained using PPO and TRPO exhibit satisfactory performance in robotics manipulation tasks.
In general, TRPO outperforms PPO in terms of SR, while PPO requires less TT. 
Moreover, both controllers demonstrate a good level of robustness against action noise.
}
\end{tcolorbox}

\section{Falsification for Defect Detection in Physical Simulators}
\label{sec.falsification_tool}

As discussed in Sec.~\ref{subsec:survey_results}, there is an urgent need to provide testing support for Isaac Sim.
However, to the best of our knowledge, a dedicated testing tool specifically designed for physical simulators is currently lacking.
To bridge this gap, we develop a falsification framework that can be seamlessly integrated with physical simulators, e.g., Isaac Sim, as well as OpenAI Gym environments (Sec.~\ref{subsec:fal_tool}).
By leveraging this framework, we compare the effectiveness of three commonly used optimization methods, i.e., random, Nelder-Mead, and dual annealing, in falsifying subject tasks (Sec.~\ref{subsec:fal_result}).

\subsection{Falsification Framework}
\label{subsec:fal_tool}

\begin{figure}
  \centering
  \includegraphics[width=.9\linewidth]{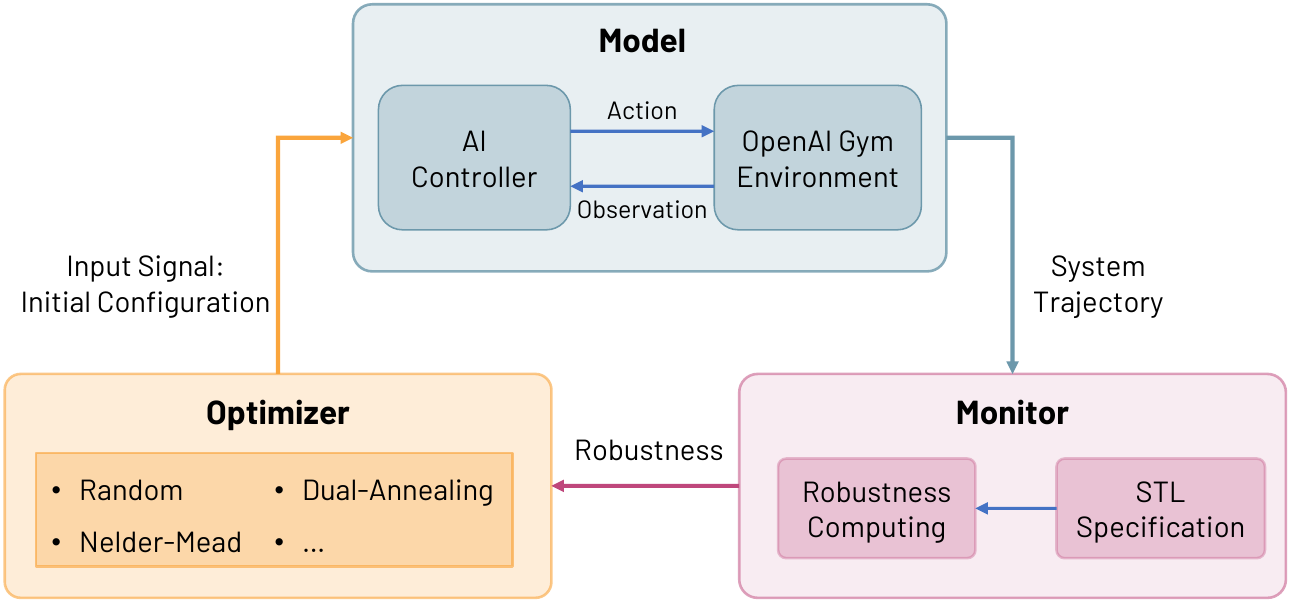}
  \caption{Structure of the proposed falsification framework.}
  \label{fig_falsification_tool}
  \vspace{-10pt}
\end{figure}

\begin{table*}[t]
\centering
\caption{Performance of different optimization methods in the falsification test (best results are highlighted in gray).}
\label{table_falsification}
\resizebox{1\textwidth}{!}{
\begin{tabular}{c|c|rrr|rrr|rrr}
\toprule
& & \multicolumn{3}{c|}{Random} & \multicolumn{3}{c|}{Nelder-Mead} & \multicolumn{3}{c}{Dual Annealing} \\
\textbf{Task} & \textbf{Controller}& \textbf{\#Suc. Fals.} & \textbf{Avg. Time} & \textbf{Avg. \#Sim} & \textbf{\#Suc. Fals.} & \textbf{Avg. Time} & \textbf{Avg. \#Sim} & \textbf{\#Suc. Fals.} & \textbf{Avg. Time} & \textbf{Avg. \#Sim}\\ \midrule
\multirow{2}{*}{PR} & PPO & \tbmgray 30 & \tbmgray 15.10 & \tbmgray 8.57 & 7 & 15.03 & 8.43 & 30 & 27.76 & 16.37 \\
                    & TRPO & 0 & - & - & 2 & 206.28 & 128.50 & \tbmgray 23 & \tbmgray 173.26 & \tbmgray 107.83\\ 
\multirow{2}{*}{CS} & PPO & 1 & 482.68 & 201 & 2 & 411.56 & 172.00 & \tbmgray 10 & \tbmgray 436.81 & \tbmgray 182.20 \\
                    & TRPO & 12	& 429.94 & 171.17 & 12 & 408.05 & 167.08 & \tbmgray 16 & \tbmgray 317.58 & \tbmgray 126.31\\ 
\multirow{2}{*}{PH} & PPO & 26 & 222.09 & 117.19 & 17 & 249.76 & 132.59 & \tbmgray 30 & \tbmgray 117.35 & \tbmgray 93.97\\
                    & TRPO & \tbmgray 27 & \tbmgray 223.78 & \tbmgray 120.41 & 13 & 189.63 & 102.00 & 26 & 220.29 & 118.27\\ 
\multirow{2}{*}{BB} & PPO & 3 & 199.22 & 101.33 & 7 & 187.56 & 96.14 & \tbmgray 9 & \tbmgray 98.85 & \tbmgray 50.22 \\
                    & TRPO & 5 & 225.87 & 111.60 & 5 & 256.52 & 126.6 & \tbmgray 6 & \tbmgray 246.68 & \tbmgray 121.67\\ 
\multirow{2}{*}{BC} & PPO & 5 & 510.06 & 195.00 & 4 & 400.12 & 153.00 & \tbmgray 6 & \tbmgray 390.68 & \tbmgray 146.50 \\
                    & TRPO & 16 & 164.51 & 56.69 & 12 & 264.35 & 90.50 & \tbmgray 25 & \tbmgray 185.03 & \tbmgray 64.32\\ 
\multirow{2}{*}{BP} & PPO & 26 & 222.09 & 117.19 & 17 & 249.76 & 132.59 & \tbmgray 30 & \tbmgray 117.35 & \tbmgray 93.97\\
                    & TRPO & \tbmgray 27 & \tbmgray 223.78 & \tbmgray 120.41 & 13 & 189.63 & 102.00 & 26 & 220.29 & 118.27\\ 
\multirow{2}{*}{DO} & PPO & 30 & 25.31 & 11.13 & 30 & 21.78 & 9.50 & \tbmgray 30 & \tbmgray 17.09 & \tbmgray 7.50\\
                    & TRPO & 30 & 58.88 & 25.83 & 30 & 56.21 & 24.30 & \tbmgray 30 & \tbmgray 54.60 & \tbmgray 23.67\\ 
\multirow{2}{*}{CP} & PPO & 16 & 121.53 & 100.00 & 2 & 158.08 & 127.00 & \tbmgray 28 & \tbmgray 62.40 & \tbmgray 50.43 \\
                    & TRPO & 21 & 1211.48 & 107.52 & 14 & 1677.16 & 99.64 & \tbmgray 30 & \tbmgray 754.18 & \tbmgray 83.13\\ \bottomrule
\end{tabular}
}
\end{table*}

Based on the proposed benchmark in Isaac Sim, we develop accordingly an extendable falsification framework to test the performance of AI controllers in physics engine-based simulations.
Fig.~\ref{fig_falsification_tool} shows the structure of the falsification framework.
It consists of three parts: the system model, the monitor, and the optimizer.

\noindent \textbf{Model:} The model is responsible for simulating the physical behavior in different robotics manipulation tasks. 
For each input signal, i.e., the initial configuration, a trained AI software controller attempts to control the manipulator to accomplish the task. 
This results in a sequence of observations, i.e., the system trajectory, that will be used in the monitor to verify the completeness of the task.
As aforementioned, in our benchmark, each task is wrapped as an Omniverse Isaac Gym environment.
To facilitate compatibility with other physical simulators, we further wrap all task environments into standard OpenAI Gym environments by using the wrapper provided in the SKRL library~\cite{serrano2022skrl}. 
This enables the developed falsification framework to be used not only with Isaac Sim but also with other physical simulators that utilize OpenAI Gym environments, e.g., Mujoco~\cite{todorov2012mujoco} or PyBullet~\cite{benelot2018}.

\noindent \textbf{Monitor:} The monitor takes the system trajectory generated from the simulation as input and computes the robustness of the AI controller's behavior against predefined STL specifications.
We use RTAMT~\cite{nivckovic2020rtamt}, a Python library for monitoring STL specifications, to compute the robustness values.
The STL specifications employed in our falsification experiments are identical to those used to define successful task completions, as detailed in Table~\ref{table_task_STL}.

\noindent \textbf{Optimizer:} The optimizer searches for worst-case scenarios where the AI controller fails to meet the STL specifications, i.e., the initial configurations resulting in a system trajectory with minimal robustness. 
Such a search process is considered an optimization problem where the objective function is the combination of the model and the monitor, which returns the corresponding robustness value for a given input signal. 
We implement three optimization algorithms, namely random, Nelder-Mead, and dual annealing, in the proposed falsification framework. 
The latter two algorithms are based on the Scipy library~\cite{2020SciPy-NMeth}. 


The proposed falsification framework provides a rigorous and systematic approach for evaluating the reliability of various AI controllers in robotics manipulation tasks. 
Based on it, we conduct a falsification experiment using our benchmark. 
The results are detailed in the next subsection. 

\subsection{Experimental Evaluations}
\label{subsec:fal_result}

To answer RQ3, i.e., how effective are different optimization methods in falsifying physics engine-based robotics manipulation tasks, we test the trained AI controllers from Sec.~\ref{sec.benchmark_result} with the proposed falsification framework. 
For each AI controller, we conduct 30 falsification trials, each consisting of a maximum of 300 task simulations.

The number of successful falsifications, the average time used, and the average number of required task simulations for successful falsifications are presented in Table~\ref{table_falsification}.
The results of our falsification test show that dual annealing outperforms other methods. 
While achieving only one less successful falsification than the random approach in the BP and PH tasks with the TRPO controller, it has a noticeable advantage in other tasks. 
In contrast, Nelder-Mead performs poorly in robotics manipulation tasks. 
A possible reason for this could be that, due to the highly nonlinear function of the robustness of STL specifications, a large number of local optima exist. 
In such cases, heuristic direct search methods like Nelder-Mead are prone to get stuck in local optima and, therefore, may fail to falsify these tasks. 

It is also worth mentioning that achieving high rewards in the training process does not necessarily indicate that the AI controller will reliably accomplish its desired task.
As mentioned in~\cite{zolfagharian2023search}, success in reward does not guarantee success in task completion.
Our tests also show that even well-trained AI controllers can still be falsifiable to state-of-the-art falsification techniques.
This highlights the necessity of incorporating falsification techniques into the testing framework of AI-enabled robotics manipulation.


\begin{tcolorbox}[size=title, colback=white]
{\textbf{Answer to RQ3:} The effectiveness of different optimization methods varies in falsifying physics engine-based robotics manipulation tasks. 
Dual annealing shows satisfactory falsification results, while Nelder-Mead has a relatively poor performance.
Considering the task-specific characteristics may be crucial when developing a highly effective falsification method for AI-enabled robotics applications.
}
\end{tcolorbox}

\section{Discussion, Future Direction, and Threats to Validity}

\noindent \textbf{Discussions.}
Based on the findings of our survey, we recognize that a versatile environment, e.g., a benchmark, is crucial for the development of AI-enabled robotics systems.
Practitioners seek an intuitive pipeline and plug-and-play solutions to streamline their efforts and avoid getting entangled in inconsequential steps. 
In response to this demand, we construct our benchmark with an emphasis on ease of use, i.e., developers can easily deploy their methods and techniques on our benchmark without the need for extensive low-level adjustments.
Moreover, our evaluation provides insights into the challenges and potential opportunities involved in developing AI software controllers with Isaac Sim. 
Notably, existing DRL algorithms may neglect the impact of the parallel running of learning environments associated with Isaac Sim. 
Further research is needed to design and develop AI methods that effectively account for and exploit this characteristic.


Although AI controllers demonstrate good performance in manipulation tasks, their reliability cannot be guaranteed according to our falsification tests. 
In the field of robotics, various traditional controllers, e.g., computed torque controller~\cite{middletone1986adaptive} and model predictive controller~\cite{camacho2013model}, have been proposed to solve manipulation tasks, which, unlike AI controllers, often provide guaranteed performance and safety assurance. 
However, these controllers are typically task-specific and rely on precise object information for motion planning, which limits their generalizability and applicability in changing environments. 
Therefore, a recent trend is to integrate AI approaches with traditional control-theoretical concepts, resulting in controllers that are flexible and adaptable with high levels of performance, safety, and reliability~\cite{luo2020balance,karoly2020deep}. 
This idea can also be applied to AI-CPSs~\cite{song2022cyber,zhang2022falsifai}, opening up new possibilities for developing trustworthy systems.

We also notice that the performance of different optimization techniques can be influenced by the properties of the objective function, which is highly dependent on the task being tested.
Therefore, taking into account the characteristics of the task is important for performing effective testing, and different methods may be necessary for different tasks.

\noindent \textbf{Future work.}
While Isaac Sim is known for its fast training speed, its ability to accurately reflect real-world behaviors requires further investigation.
For applying AI controllers in real-world industrial applications, an analysis of the simulation-to-reality gap is critical.
Therefore, one of our future research directions is to investigate whether Isaac Sim can help overcome this gap.
Another possible direction is to incorporate more state-of-the-art optimization methods, such as those available in Breach~\cite{donze2010breach} or S-Taliro~\cite{Annpureddy-et-al2011} libraries, into our falsification framework.
This would increase the versatility of our framework, making it useful not only for robotics manipulation tasks but also for other types of AI-CPSs. 

\noindent \textbf{Threats to validity.}
To address the \textit{construct validity} concern, we recognize that the  evaluation metrics may not fully capture the performance of AI controllers. 
Therefore, we adopt multiple evaluation metrics that cover various perspectives of the task to measure and analyze the performance of AI-enabled robotics manipulation.
In terms of \textit{internal validity}, one potential threat is that the behavior of the AI controller can differ when using different training parameters. 
To mitigate this threat, we test different sets of parameters and choose the one that results in the best performance in our evaluation.
Regarding the \textit{external validity}, we acknowledge that our evaluation results may not be generalized to other robotics manipulation tasks. 
To address this concern, we include a diverse set of tasks with unique challenges and requirements in our benchmark to consider a wide range of manipulation scenarios.

\section{Related Work}


\noindent{\textbf{Benchmark of CPS and AI-CPS.}}
CPSs are highly integrated systems that collaborate diverse disciplines, e.g., mechanical, electrical and particularly software engineering, to tackle challenges in real-world applications. 
Ernst et al.~\cite{ernst2020arch} provided a benchmark of traditional CPSs and compared the performance of various testing tools on these systems. 
However, this benchmarks only focuses on traditional CPSs instead of AI-CPSs.
Song et al.~\cite{song2022cyber} proposed a first benchmark for AI-CPSs with nine tasks from different domains. 
However, this benchmark is built on MATLAB, which relies on accurate mathematical models to describe the system behaviors. 
Although this model-based simulation is capable of design and prototyping, we consider a physical simulator can better reflect the complexity of industrial-level operations. 
In addition, other existing literature~\cite{duan2016benchmarking, johnson2020arch, braganza2007neural} includes AI-CPSs, but most of them are simplified systems or game scenarios such as \emph{Cart-Pole} and \emph{Inverted Pendulum} that are not appropriate for industrial-level applications.

\noindent{\textbf{Benchmark of robotics.}} 
In the field of robotics, multiple benchmarks have been proposed for general robotics applications, e.g.,~\cite{calli2015benchmarking,fan2018surreal,ahn2020robel}. 
However, these benchmarks often have limitations, as they either focus on providing different robot, object, or sensor models~\cite{zhu2020robosuite,mittal2023orbit} or are designed only for specific topics of robotics control, such as manipulating deformable objects~\cite{chatzilygeroudis2020benchmark} or real-time robotics~\cite{bakhshalipour2022rtrbench}. 
Moreover, these benchmarks often ignore important components of the software development lifecycle, e.g. testing.
Considering these limitations, a unified benchmark that covers a broad range of tasks and effectively supports the software development lifecycle is desirable.
By proposing a benchmark based on Isaac Sim, we aim to address this need and take a first step towards building a development platform for AI-enabled robotics applications.


\noindent{\textbf{AI-CPS testing.}}
Testing is a non-trivial topic in CPSs as it provides the quality assurance to deploy reliable and robust CPSs in safety-critical applications. 
Such that, there is an increasing trend of research devoting in this direction~\cite{zhang2022falsifai, zolfagharian2023search, huang2017safety, xiang2018output}. 
Zolfagharian et al.~\cite{zolfagharian2023search} proposed a search-based testing approach that leverages a genetic algorithm to generate testing cases for DRL agents.
Zhang et al.~\cite{zhang2022falsifai} leveraged the temporal behaviors of DNN controllers and introduced a falsification framework for AI-CPSs. 
In~\cite{thibeault2021psy}, a Python-based falsification toolbox for CPSs is presented.
However, like other MATLAB falsification tools, it still requires an accurate system model defined as, e.g., ordinary differential equations.
Moreover, it is still unclear the effectiveness of existing testing approaches on modern physical simulators. 
Our work develops a Python-based extendable falsification framework with various optimization methods for physical simulators as well as OpenAI Gym environments. 
We believe that our framework can greatly enhance the flexibility of conducting testing for AI-CPS practitioners and motivate further research along this direction. 

\section{Conclusion}
By using Isaac Sim, this paper presents a public industrial benchmark of AI-enabled robotics manipulation that includes eight representative manipulation tasks.
Multiple AI software controllers are trained with various DRL algorithms, and an evaluation of their performance is conducted.
The results show that AI controllers are able to successfully complete the tasks with satisfactory performance and a good level of robustness against action noise.
To further test the AI controllers, we also develop the first Python-based falsification framework that is compatible with physics engine-based simulators and OpenAI Gym environments.
Three different optimization methods are employed to falsify AI controllers in robotics manipulation tasks.
The results of the falsification test reveal the effectiveness of state-of-the-art falsification techniques in identifying system defects, making them useful for analyzing the reliability of AI controllers.
Our work establishes a foundation as well as a systematic pipeline for evaluating and testing AI-enabled robotics systems with modern physical simulators, which is an important step towards understanding and developing trustworthy AI-CPSs for critical real-world domains.

\bibliographystyle{ACM-Reference-Format}
\bibliography{ref}







\end{document}